# Co-doping Er plus V or Er plus Nb into CaWO$_4$

Chen Yang[1,2], and Robert J. Cava[1*]

[1] Department of Chemistry, Princeton University, Princeton, New Jersey, 08544

[2] Department of Electrical and Computer Engineering, Princeton University, Princeton, New Jersey, 08544

*corresponding author's email: cy11@princeton.edu

**Abstract**

Er$^{3+}$ plus V$^{5+}$, and Er$^{3+}$ plus Nb$^{5+}$ co-doped CaWO$_4$, formulas Ca$_{1-x}$Er$_x$W$_{1-x}$M$_x$O$_4$, were synthesized in air by a conventional solid-state method. A color change from white to pink was observed in the final products. An equal fraction of dopants was employed to obtain charge neutrality, and the limits of the solubility for our conditions are lower than $x$=0.15. The magnetic susceptibility data shows that that the magnetic coupling becomes increasingly antiferromagnetic with increasing Er$^{3+}$ content. The Curie-Weiss fit and isothermal magnetization imply that different degrees of spin-orbit coupling appear to be present in the two doping systems. No transitions were observed in the heat capacity data above 0.4 K.

**Introduction**

CaWO$_4$, a well-known scheelite, is a widely used material in LEDs. Many groups have implanted rare-earth ions into this material to manipulate its photoluminescence (PL) or optical thermometry [1–7]. Er–implanted CaWO$_4$ has also recently been studied as a promising host material for quantum communications, motivating the current study. This application benefits from properties such as the tetragonal structure, the lack of a center of symmetry on the Ca$^{2+}$ site, the low rare-earth-based background, and the zero nuclear spin of CaWO$_4$ [8,9]. This paper describes the successful co-doping of Er$^{3+}$ plus V$^{5+}$ or Er$^{3+}$ plus Nb$^{5+}$ into CaWO$_4$ in air to maintain charge neutrality. Although putting non-zero nuclear spin ions into the system defeated our initial purpose of improving the host material performance for quantum communications, the result offered a new idea to alter the electronic structure that may improve the host material performance. The successful co-doping also provides a new method to tune the luminescence of CaWO$_4$ for other optical applications.

**Experimental**

The synthesis was via traditional solid-state reaction. The starting materials were CaCO$_3$ (Alfa Aesar, 99.9999%), WO$_3$ (Alfa Aesar, 99.8%), Er$_2$O$_3$ (Thermo Scientific, 99.99%), V$_2$O$_5$ (Johnson Matthey Electronics, 99.9%), and Nb$_2$O$_5$ (Thermo Scientific, 99.9985%). To ensure that all the reagents were free of moisture, all reagents other than $Er_2O_3$ were pre-dried overnight at 120°C, and $Er_2O_3$ was dried at 900°C overnight. After weighing the reactants in a molar ratio, the combination was mixed in an agate mortar and pestle. The samples were then heated in air (Sentro Tech Corp. ST-1600C-445 High-Temperature Box Furnace) to 900° C at 180° C/min and held for 8 hours. Subsequently, the temperature was set to 1200° C with the same heating rate and held for 48 hours with intermediate grindings until achieving a single-phase product. The powder x-ray diffraction (PXRD) patterns used to characterize the products were collected by a Brucker D8 FOCUS diffractometer with Cu Kα radiation ($\lambda_{K\alpha}$= 1.5406 Å) and refined by Le Bail and Rietveld fitting [10,11]. Magnetic susceptibility and $^3$He heat capacity measurements were carried out in a Quantum Design Dynacool Physical Property Measurement System (PPMS).

**Results and Discussion**

The Stoichiometric and both $x$=0.1 PXRD patterns were refined by the Rietveld method in the GSAS II software package to verify that our nominal chemical formulas were correct (**Fig. 1**). $CaWO_4$ is reported to be a tetragonal scheelite, with lattice parameters a = 5.243Å and c=11.376Å [12]. Both sets of doped samples were isostructural with stoichiometric $CaWO_4$ with no impurity peaks detected for $x$=0.05 and 0.1. For Er plus V, $x$=0.15 peaks were broadened and impurity peaks belonging to $ErVO_4$ started to emerge. For Er plus Nb, clear impurity peaks from niobium oxide and doublets formed, suggesting a lowering of the symmetry of the bulk material. Thus, our evidence shows that the saturation limit for the tetragonal symmetry scheelite solid solution for our synthetic conditions appears to lie between $x$=0.10 and $x$=0.15.

It has been reported many times that a low concentration of $Er^{3+}$ or other rare-earth ions can induce a color change in $CaWO_4$. This is consistent with the fact that our stoichiometric sample is white (**Fig 1a**) while our doped $CaWO_4$ samples are pink (**Figs. 1b and c**). The samples display different shades of pink [13–15] due to the influence of the dopant on the $W^{6+}$ site.

The PXRD patterns (**Fig. 2**) show a consistent peak shift for both doping sets. However, in the PXRD patterns, we observed inconsistent peak shift directions (**Fig. 3**) plotting Le Bail fitted lattice parameters of all doped samples. We noticed that lattice parameters *a* and *c* change differently for the V and Nb substitutions. $V^{5+}$ and $Nb^{5+}$ in a tetrahedron are expected to have an ionic radius of 0.3 Å and 0.48 Å, respectively [16]. It is thus surprising that the *c*- axis contraction caused by $V^{5+}$ doping is less than that caused by $Nb^{5+}$.

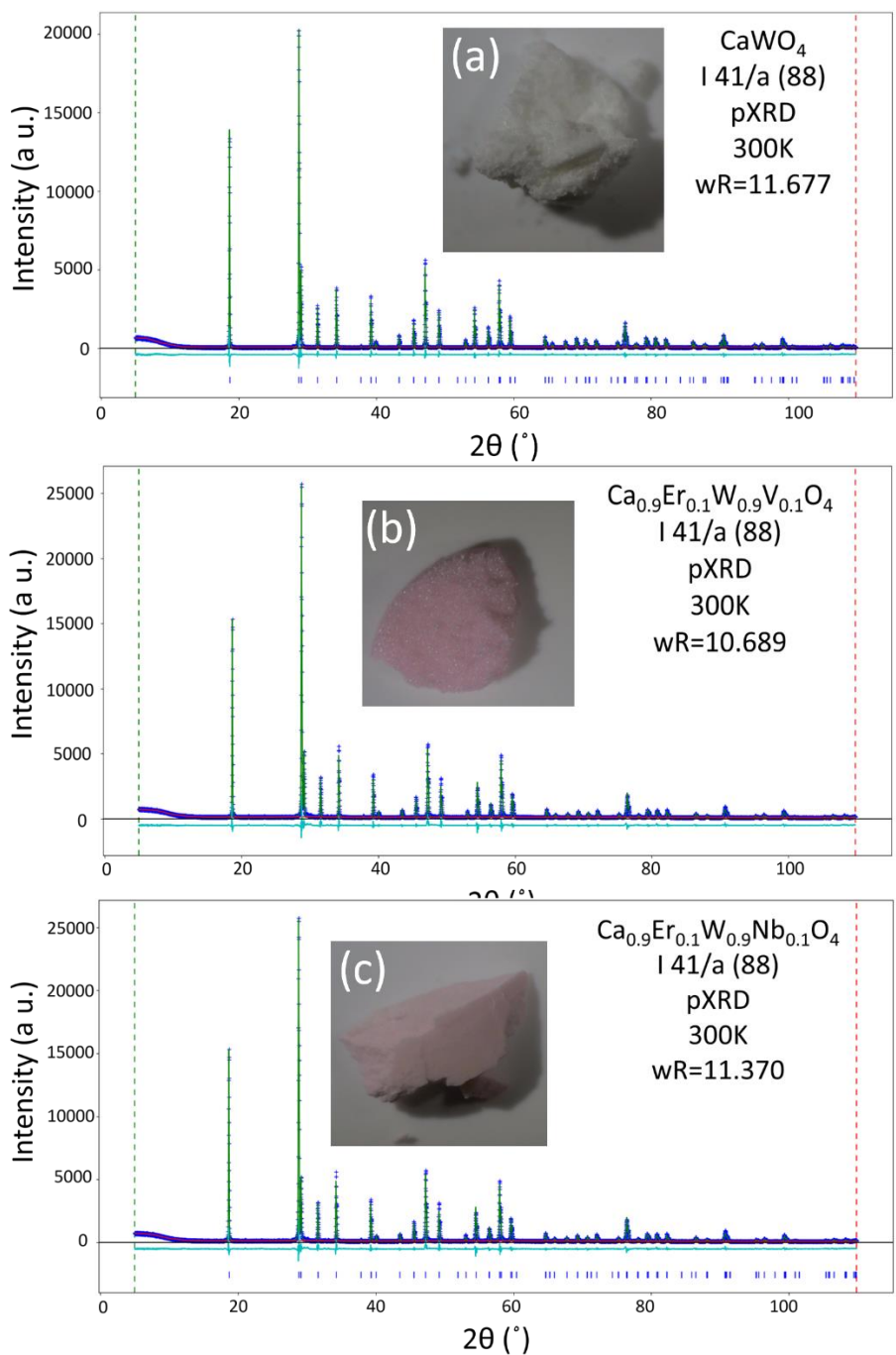

**Figure 1. The refined (Reitveld method) PXRD patterns and sample photos for $Ca_{1-x}Er_xW_{1-x}M_xO_4$.** (a) $CaWO_4$, x = 0 (b) M = V, x = 0.1 (c) M = Nb, x = 0.1 (Crystallographic parameters are in **Table S1**.)

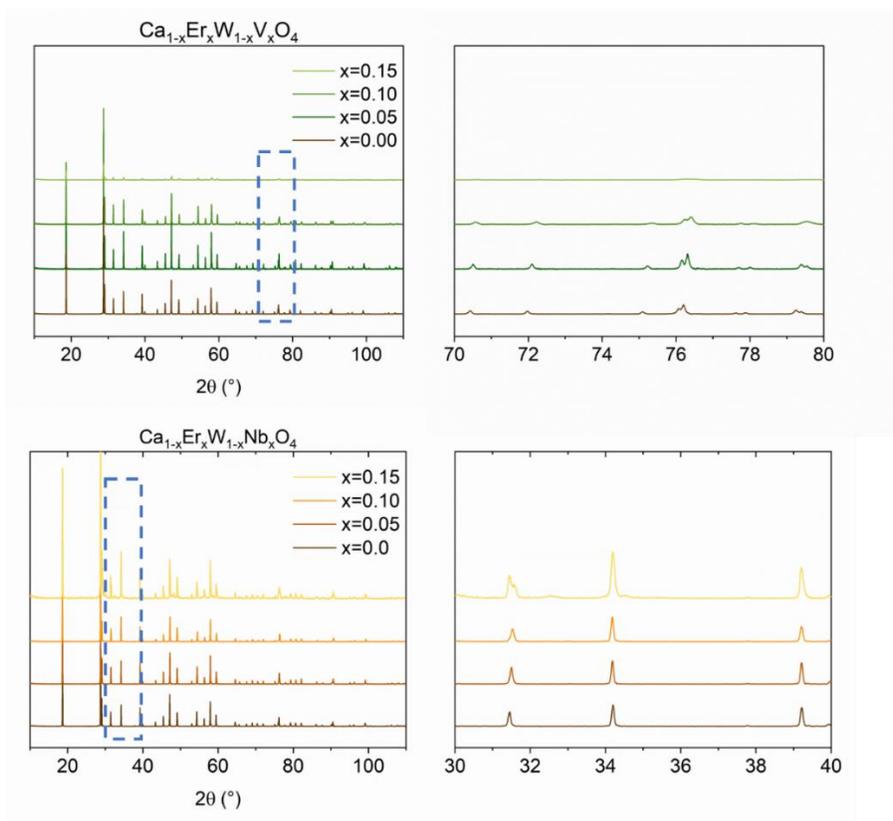

**Figure 2. All sample PXRD patterns.** The expanded view highlighted by the blue-dashed frames on the left is shown on the right.

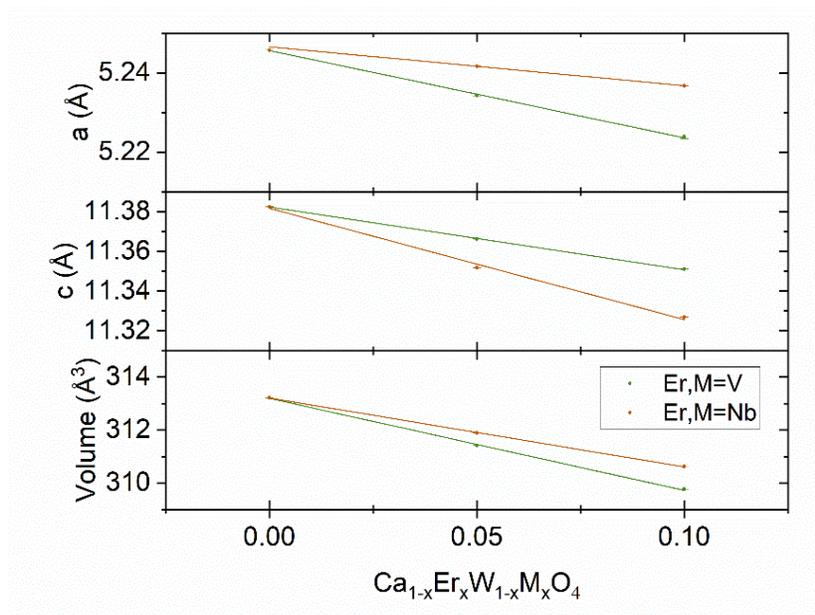

**Figure 3. Lattice parameter vs. composition in Ca₁₋ₓErₓW₁₋ₓMₓO₄.** The Le Bail fitted lattice parameters follow Vegard's Law. The error bars are smaller than the data points.

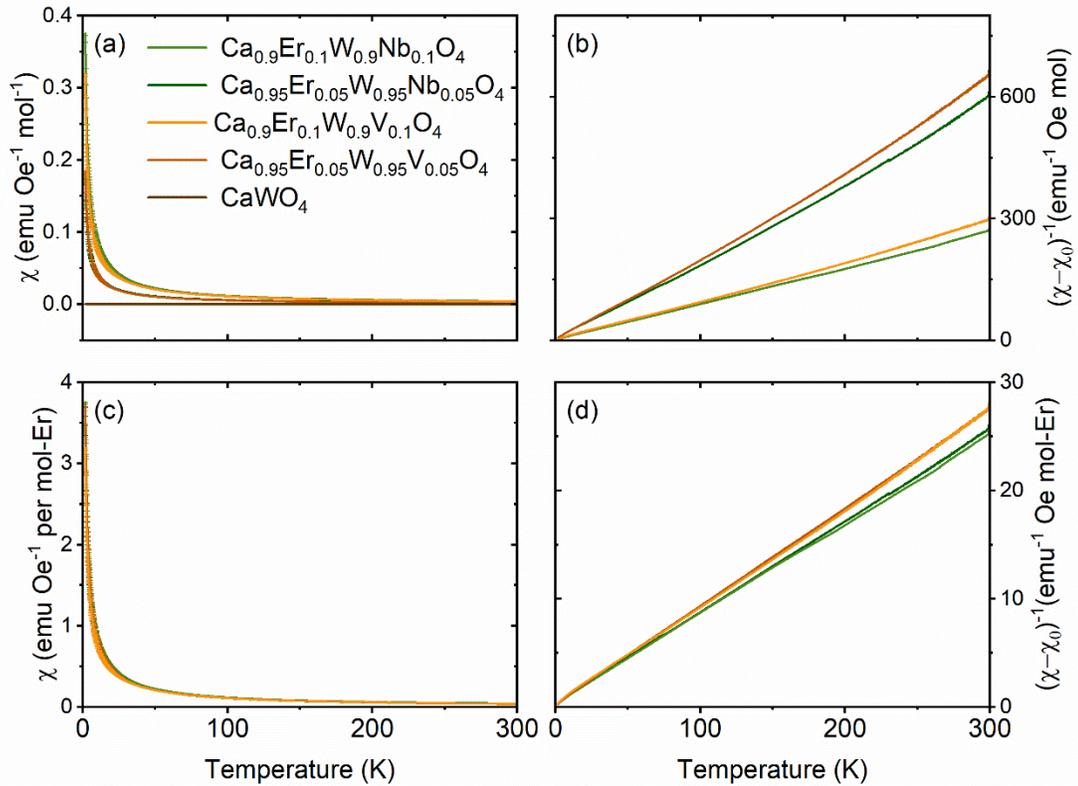

**Figure 4. The temperature-dependent magnetic susceptibility (a)(b) per mol and (c)(d) per mol-Er.**

In **Fig. 4**, the temperature-dependent magnetic susceptibility is shown per mole formula unit (4a, b) and per mole Er (4c, d). In Figs. 4b and 4a, it is evident that the magnetic moment segregated into two groups. The difference is small but appears to be consistent. Given that the only magnetic ion in the samples is Er, it is surmised that the effective moments per Er in the two doping sets differ. The isothermal magnetization data (**Fig. 5**) shows that the Er in the W/V-disordered system has a bigger moment than Er in the W/Nb-disordered system in $CaWO_4$. We postulate that the difference can be attributed to differences in the $WO_4$ tetrahedra caused by the co-dopants, which would yield different crystal field environments. We performed the Curie-Weiss fitting on Figure 4d in the 50-100K temperature range where the fitted $\mu_{eff}$ is closer to the theoretical value. However, considering the complexity of rare-earth ion magnetism, we fitted the $1/\chi$ per Er data in the 2-5K temperature range as well (**Table 1**). In **Fig S1**, we plotted the curie temperature verses the fitted lattice parameters. Since we observe no correlation between magnetism and lattice parameters, we can safely make the conclusion that the magnetism is independent of the lattice size in this system.

**Table 1. Curie-Weiss fitted parameters for all the doped samples in two different fitting ranges.**

| $Ca_{1-x}Er_xW_{1-x}M_xO_4$ | Temperature (K) | $\theta_{cw}$ (K) | $\mu_{eff}$ ($\mu_B$) |
|---|---|---|---|
| M=V, x=0.05 | 2-5 K | -0.443(2) | 7.936(4) |
|  | 50-100 K | -6.564(9) | 9.876(1) |
| M=V, x=0.1 | 2-5 K | -0.548(1) | 8.316(4) |
|  | 50-100 K | -3.450(7) | 9.745(7) |
| M=Nb, x=0.05 | 2-5K | -0.528(4) | 8.214(1) |
|  | 50-100 K | -4.002(3) | 9.448(1) |
| M=Nb, x=0.1 | 2-5K | -0.542(8) | 7.668(1) |
|  | 50-100 K | -5.616(8) | 9.592(1) |

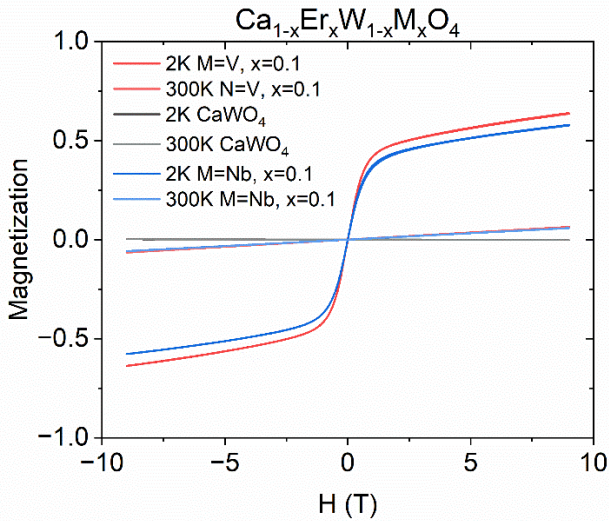

**Figure 5. The isothermal magnetization for all samples at 5 K and 300 K.**

Finally, we measured the temperature-dependent heat capacity to 0.4 K under 0 applied magnetic field. The fact that there is no very low temperature anomaly seen for undoped $CaWO_4$ indicates that there is no nuclear spin ordering present due to unpaired nuclear spins on those elements. No complete transition was observed in *x*=0.05 or *x*=0.1 (**Fig. 6a**). The maximum C/T obtained was different for different Er concentrations. In addition, we measured capacity under zero magnetic field for the different dopants at the same doping concentration (**Fig. 6b**). The transition was incomplete, and we cannot pinpoint the transition temperature for either case.

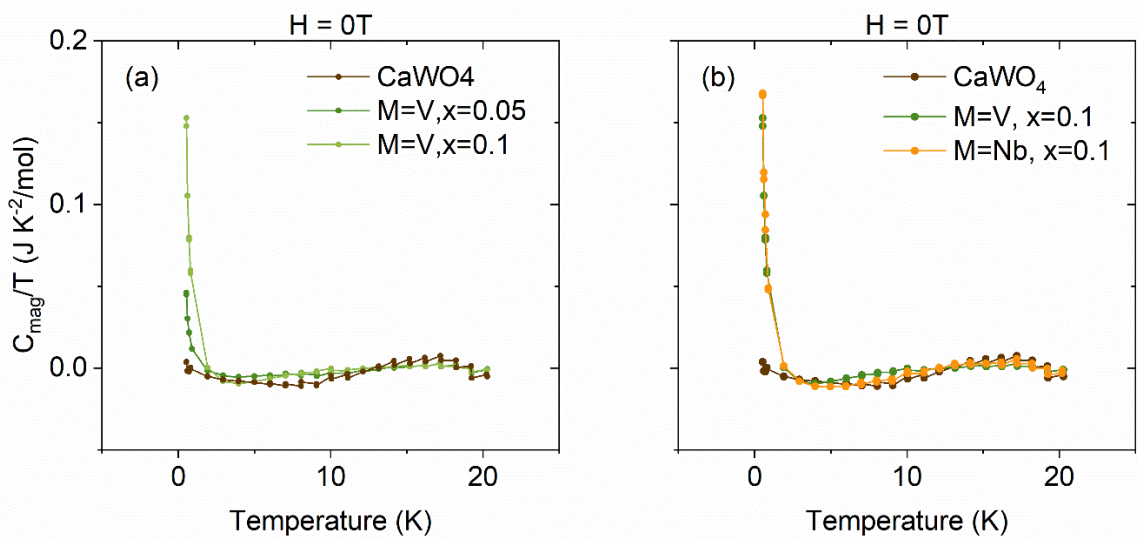

**Figure 6.** The $^3$He Heat capacity data (a) comparing different x for the same co-dopant (b) comparing different co-dopants with the same x.

## Conclusion

We successfully synthesized Er plus V or Er plus Nb co-doped $CaWO_4$ and observe a clear color change. Both kinds of samples are pink though the depth of the color differs. The PXRD refinement follows Vegard's law, but the gradients differ for the *a* and *c* lattice parameters. The temperature-dependent magnetic susceptibility showed that $Er^{3+}$ experiences different spin-orbit coupling in the V and Nb co-doped samples. These differences may be a reflection of differences in the co-dopant distribution. The Curie-Weiss temperature is independent of the lattice size. No complete transitions were observed in He-3 heat capacity data, but it can be surmised that the transition temperature will not be shifted substantially by co-dopant type or Er concentration. In future work, photoluminescence experiments or more optical investigation can be compared with only Er-doped $CaWO_4$, and determination of the local structures may be of interest.

## Acknowledgments

We would like to thank the group of Jeff D. Thompson for inspiration on $CaWO_4$. This paper is based upon work supported by the U.S. Department of Energy, Office of Science, National Quantum Information Science Research Centers, Co-design Center for Quantum Advantage (C2QA) under contract number DE-SC0012704.